\documentclass[english,aps,prper,reprint,showpacs,titlepage,longbibliography]{revtex4-2}   

\usepackage[T1]{fontenc}	
\usepackage[latin9]{inputenc}	
\usepackage{geometry}		
\usepackage{graphicx}
\usepackage[above,below]{placeins}	
\usepackage{times}

\usepackage{hyperref}  
\hypersetup{colorlinks=true,urlcolor=blue,citecolor=blue,linkcolor=blue}   
\usepackage{enumitem}          
\setlist{nosep}                 

\DeclareRobustCommand{\okina}{%
  \raisebox{\dimexpr\fontcharht\font`A-\height}{%
    \scalebox{0.8}{`}%
  }}%

\newcommand{\quotate}[2]{\begin{itemize} \item[]\reviewerA{``}#1\reviewerA{''} (#2) \end{itemize}}
\newcommand{\Choice}{{\sc{Choice}}}
\newcommand{\choice}{{\sc{choice}}}

\newcommand{\domain}{{\sc{domain}}}

\newcommand{\system}{{\sc{system}}}

\newcommand{\managerial}{{\sc{managerial}}}

\newcommand{\NOS}{{\sc{nature of science}}}
\newcommand{\Execution}{{\sc{Execution}}}
\newcommand{\execution}{{\sc{execution}}}
\newcommand{\Deliverables}{{\sc{Deliverables}}}
\newcommand{\deliverables}{{\sc{deliverables}}}
\newcommand{\Understanding}{{\sc{New Understanding}}}
\newcommand{\understanding}{{\sc{new understanding}}}

\newcommand{\reviewerA}[1]{#1}
\newcommand{\reviewerB}[1]{#1}

\usepackage{xcolor}
\newcommand{\confidentialreview}[2]{{{#1}}} 

\begin{document}

\begin{titlepage}

\title{Student ownership and understanding of multi-week final projects}

\author{Ira Ch\'e Lassen}
\affiliation{Department of Physics and Astronomy, Western Washington University, Bellingham, WA 98225, USA}
\affiliation{Fairhaven College of Interdisciplinary Studies, Western Washington University, Bellingham, WA 98225, USA}

\author{Acacia Arielle-Evans}
\affiliation{Department of Geology, Western Washington University, Bellingham, WA 98225, USA}

\author{Laura R\'ios}
\affiliation{Department of Physics, California Polytechnic State University, San Luis Obispo, CA, USA 93407}

\author{H. J. Lewandowski}
\affiliation{Department of Physics, University of Colorado Boulder, Boulder, CO 80309, USA}
\affiliation{JILA, National Institute of Standards and Technology and University of Colorado Boulder, Boulder, CO 80309, USA}

\author{Dimitri R. Dounas-Frazer}
\affiliation{Department of Physics and Astronomy, Western Washington University, Bellingham, WA 98225, USA}
\affiliation{SMATE, Western Washington University, Bellingham, WA 98225, USA}


  \begin{abstract}
    
    National calls to transform laboratory courses by making them more discovery based can be met by engaging students in multi-week final projects. One plausible outcome of this approach is that students may feel ownership of their projects. We define ownership as a dynamic relationship between students and their projects characterized by three student-project interactions that evolve over three project phases. Student-project interactions include students' contributions to, emotional responses to, and new understanding of the project. Phases include choosing the topic and team, carrying out the research, and creating and presenting end-of-project deliverables. Drawing on interviews with students collected as part of a multi-year, multi-institutional study, this paper will elaborate on the  evolution of students' own new knowledge about the project across different project phases. Throughout our work, we compare the type of ownership that manifests when students work on projects in a lab course to the type of ownership gardeners feel when tending to plots in a community garden. We end with recommendations for instructors wishing to foster project ownership in their lab courses.

    
   \clearpage
  \end{abstract}

  \maketitle
\end{titlepage}

\section{Introduction and background}

According to the American Physical Society, providing undergraduate students with increased opportunities to engage in research-style activities is a national priority~\cite{APS2014}. Meanwhile, project-based approaches to physics lab instruction have been gaining popularity~\cite{Feder2017}. In a 2012 report, the President's Council of Advisors on Science and Technology (PCAST) suggested that standard laboratory courses could be replaced with discovery-based research courses~\cite{PCAST2012}. The PCAST report framed student ownership of projects as a plausible desirable outcome of such a shift~\cite{PCAST2012}. Along these lines, our team has been studying student ownership of projects in upper-level physics lab courses that include multi-week student-driven experiments.

We interpret the type of ownership that manifests among students during project portions of lab courses as similar to the ownership that manifests among gardeners in a community garden. In a community garden, a gardener submits an application to temporarily tend a plot of land, and they sometimes have access to communal tools (e.g., shovel, wheelbarrow, and hose). Although the land they tend and the tools they use may not be their own private property, the gardener can choose which seeds to plant, which fertilizers or pesticides to use, and so on. Further, it would be a violation of community norms for other gardeners to add or remove plants from the plot. In this sense, the gardener may view the plot as `their own.' A student working on a final project in a physics lab may feel a similar type of ownership with their project. They have temporary access to lab space in which they have some control over an apparatus.
Although the lab space and equipment are not their own private property, classroom norms establish boundaries that prohibit other people from making changes to the apparatus. Thus, like the gardener and the garden plot, the student may view the project as `their own.' It is this type of ownership that we explore.

Our ideas about ownership have been influenced in part by work from beyond the science education literature~\cite{Savery1996, Wiley2009} as well as work in physics~\cite{Enghag2006, Milner-Bolotin2001} and biology~\cite{Hanauer2012, Hanauer2014} education. For a detailed summary of literature on ownership, see Ref.~\cite{Dounas-Frazer2017}. Here, we focus on three patterns in that literature.

First, ownership is commonly modeled as dynamic. For example, Milner-Bolotin (2001) found that students' sense of project ownership, as measured by the Ownership Measurement Questionnaire (OMQ), was low in the middle of a project, when students were navigating challenges with their projects, and increased toward the end of the project, when \reviewerA{``}students started seeing the fruits of their hard work.\reviewerA{''} (p.139) \cite{Milner-Bolotin2001}. Similarly, Enghag (2006) argued that ownership corresponds to students' level of choice and control during three project stages: choosing, performing, and presenting the project~\cite{Enghag2006}. These features of ownership are consistent with Papert and Harel's (1991) interpretation of constructionism~\cite{Papert1991}, in which the acts of creation followed by presentation play key roles in the learning process.
On smaller timescales, frequent cycles of frustration and success contribute to students' sense of ownership,~\cite{Dounas-Frazer2017} and may also support students to feel proud of their work~\cite{Little2015}.
In our model, just as a gardener's sense of ownership evolves throughout planting and harvesting seasons, a student's sense of ownership evolves across \reviewerA{three} project phases: choice of topic, execution of methods, and creation of deliverables~\cite{Dounas-Frazer2019}.

Second, ownership frameworks also typically include students' ability to make choices about what they learn and how they learn it. For example, Savery (1996) identified control of the learning environment as an affective indicator of ownership, related to students' intrinsic motivation and beliefs about their own competence~\cite{Savery1996}. Similarly, Wiley's (2009) definition of ownership includes the right and responsibility of students to determine the fate of their learning~\cite{Wiley2009}.
In  physics education, Milner-Bolotin and Enghag identify, respectively, that students' feelings of control and responsibility~\cite{Milner-Bolotin2001} and their influence and impact on the learning environment~\cite{Enghag2006} are features of project ownership. In  biology education, Hanauer et al.\ (2012) framed project ownership as a \reviewerA{``}complex interaction between the student and the educational environment\reviewerA{''} (p.~379) and argued that feelings of ownership are fostered when students' are able to make decisions about research questions and methods~\cite{Hanauer2012}.
Similarly, a gardener's relationship with their garden involves the gardener making contributions to the plants they grow, and we model ownership as a student-project relationship in which a student makes contributions to the experiment they carry out~\cite{Dounas-Frazer2019}.

Third, students' sense of ownership is related to their pursuit and generation of knowledge. For example, Savery defined knowledge construction, motivation, and attitude toward learning as factors that comprise ownership~\cite{Savery1996}. Milner-Bolotin found that students' ownership of physics projects was coupled to their intrinsic motivation to master a new topic~\cite{Milner-Bolotin2001}. Building on these studies, Enghag found that physics student's motivation to work on mini-projects was positively impacted by their sense of ownership, which in turn had a positive impact on their holistic understanding of physics~\cite{Enghag2006}. More recently, \confidentialreview{Dounas-Frazer and Lewandowski (2019)}{a previous study} used student responses to the Project Ownership Survey (POS)~\cite{Hanauer2014} and the Colorado Learning Attitudes about Science Survey for Experimental Physics (E-CLASS)~\cite{Zwickl2014,Wilcox2018} to show a positive correlation between students' sense of project ownership and their attitudes about experimental physics~\cite{Dounas-Frazer2018a}. 

\reviewerA{In this paper, we explore one aspect of our project ownership model; how students' development of their own new understanding of the project evolves throughout all three project phases.} Our research is guided by the following questions: When discussing their own new knowledge about the project,
\begin{enumerate}
\item Which types of knowledge did participants describe developing, and
\item During which phases did they describe developing each type of knowledge?
\end{enumerate}
To answer these questions, we analyzed 15 interviews in which students discussed their experiences working on multi-week projects in upper-level physics labs.

\section{Context, theory, and methods}

Participants were \reviewerB{undergraduate} students enrolled in, or instructors teaching, upper-division physics lab courses at one of five universities: a selective, predominantly white baccalaureate college; two inclusive, Hispanic-serving master's universities; and two selective, predominantly white doctoral universities~\footnote{The 2018 edition of the Carnegie Classification System classifies institutions of higher education according to degree type and selectivity of admissions processes, and the Hispanic Association of Colleges and Universities maintains a database of Hispanic-Serving Institutions.} In each lab course, students were required to work in groups of 2--4 to complete projects that lasted 4--7 weeks. Project topics included acoustic levitation, thermal lensing, and other topics. We administered E-CLASS and POS to students, collected weekly project reflections from students, collected student notebooks and reports, and conducted post-project interviews with students and instructors. In total, 87 students and 4 instructors agreed to participate in one or more aspect of the study. Analyses of subsets of these data have been reported elsewhere~\cite{Dounas-Frazer2019,Dounas-Frazer2018a}. \reviewerA{Of the 87 student participants, only 15 agreed to participate in post-project interviews. Those interviews are the focus of this work.}


Interviews were conducted remotely prior to the start of the COVID-19 pandemic. They lasted 60--80 minutes, for a total of about 17 hours. Interview protocols were developed by \confidentialreview{DRDF, LR, and HJL}{authors 3, 4, and 5} and conducted by \confidentialreview{DRDF}{author 5} using the life grid method. This method, which originated in medical sociology~\cite{Blane1996} and has been adapted for use in education research~\cite{Abbas2013}, involves the interviewer and interviewee collaboratively filling out a grid for which rows correspond to intervals of time and columns to kinds of events related to the phenomenon being studied. The method's purpose is to reduce recall bias in participants and facilitate collaborative interviews. In our application, the life grid was used to probe student interactions with the experiment, their group members, and instructors during each week of the project. A detailed description of our approach is provided in Ref.~\cite{Rowland2019}.

Data analysis was guided by a model initially developed by \confidentialreview{DRDF, LR, and HJL}{authors 3, 4, and 5}~\cite{Dounas-Frazer2019}. That model was informed by preliminary analysis of the larger data set affiliated with the present work. Specifically, based on instructor and student responses to interview questions that asked interviewees to describe what it meant to them for students to feel ownership of a project, ownership was modeled as a relationship between students and projects that consists of three types of student-project interactions and evolves across three sequential project phases. Student-project interactions consist of students' own contributions to, emotional responses to, and new knowledge about the project. Project phases refer to the periods of time during which students choose or propose projects and teams, carry out research, and synthesize their notes, results, and new knowledge to create summative reports and presentations~\cite{Dounas-Frazer2019}.

The goal of the present work is to characterize one aspect of project ownership---namely, students' own new understanding of the project---across all three project phases. We refined the definition of the synthesis phase, which we now refer to as the deliverables phase. We did so because interviewees almost always described engaging in synthesis of notes, results, and new knowledge during the execution phase. When analyzing data, we used the ownership-as-relationship model~\cite{Dounas-Frazer2019} as an \emph{a priori} coding scheme with the following operational definitions:
\begin{itemize}
    \item \Understanding: student described changes in their perspective, knowledge, or abilities.
    \item \Choice: student described process of selecting topic, team, or initial project goals. 
    \item \Execution: student described carrying out the experiment, including apparatus design and data analysis.
    \item \Deliverables: student described creating or modifying content for end-of-project deliverables.
\end{itemize}

We used a multi-pass approach to data analysis. The first pass was conducted via remote collaboration due to the COVID-19 pandemic. \confidentialreview{ICL, AAE, and DRDF}{Authors 1, 2, and 5} collaboratively analyzed data in consultation with \confidentialreview{LR and HJL}{authors 3 and 4}.  During the first pass, we applied the \emph{a priori} coding scheme to all 15 interview transcripts. \confidentialreview{ICL and AAE}{authors 1 and 2} acted as a single rater, and \confidentialreview{DRDF}{author 5} as a second rater who reviewed coded excerpts and flagged excerpts that might not fit the definition of the code that was assigned to them. \confidentialreview{DRDF}{Author 5} flagged fewer than 6\% of about 700 excerpts coded according to phase and fewer than 12\% of about 900 excerpts coded according to interaction. Discrepancies were resolved through discussion. Throughout this pass, we regularly discussed our methods and interpretations with researchers and other stakeholders in science education~\footnote{\confidentialreview{Participants in weekly dialogues include L.\ Dahlberg, E.\ Duffy, and N.\ Stephenson; G.\ Quan and A.\ Little; R.\ Barthelemy and A.\ Knaub; J.\ Behrman and B.\ Zamarripa Roman; and G.\ McGrew and L.\ Osadchuk.}{Redacted for confidential review.}} in order to incorporate dialogue into our process for generating claims~\cite{HillCollins1989}. Early in this process, \confidentialreview{ICL and AAE}{authors 1 and 2} recognized that the creation of deliverables should be the defining characteristic of the third project phase.

The second pass was conducted via in-person collaboration. To reduce risk of transmission of COVID-19, \confidentialreview{ICL and DRDF}{authors 1 and 5} worked in a ventilated room, wore masks, and ceased work after two hours of continuous collaboration. During the second pass, \confidentialreview{ICL and DRDF}{authors 1 and 5} collaboratively implemented emergent thematic analysis for the subset of about 220 excerpts initially coded as instances of \understanding.  Emergent themes each corresponded to a type of knowledge, including the four following types:
\begin{itemize}
\item \domain: knowledge about the concepts, theories, principles, and phenomena that underlie the project.
\item \system: knowledge about the purpose of apparatus, equipment, software, and components---including how they are supposed to work versus actually work.
\item \managerial: knowledge about when and how to set or revise goals, divide labor, manage time, document progress, solicit help, and communicate results.
\item \NOS: knowledge about what experimental physics entails.
\end{itemize}

In the context of a community garden, a gardener's sense of ownership might take the form of \understanding\ about their garden plot: learning about the effect of minerals in their soil (\domain), how to use new gardening tools (\system), or what it feels like to be beholden to the weather (\NOS). \managerial\ knowledge might manifest when multiple gardeners collaboratively tend a plot.



\section{Results\label{sec:results}}
We organize the results of our emergent thematic analysis of students' \understanding\  into three sections by project phase: \choice, \execution, and \deliverables. In each section, we present one or more representative excerpts of interview transcripts from students whose self-selected pseudonyms are Jordan (she/her), Francisco (he/him), Carlos (he/him), Alyssa (he/him), and Olivia (she/her). During interviews, each of these participants described feeling high levels of ownership of their project.

In total, nine participants each discussed new understanding in the \choice\  phase once or twice during their interviews.
Of these, four described gaining \domain\  knowledge, and five students described gaining \managerial\  knowledge.
For example, when summarizing what she had learned from her project, Jordan said,

\quotate{There should be one thing in the project that draws strongly on at least one of every member's---what they might see as one of their key strengths---because I think that feeling like you're contributing is really important. But then, also, that there's at least one part of the project that allows each group member to expand their knowledge. So, I guess, in terms of learning, I feel like, in the future, I'm going to be---I'm a lot more conscious now about the ways in which work is balanced in a project, and the ways in which doing preliminary work in a project can determine---um, it can be really beneficial to the long term success of the project. I think that, actually, the stage before the preliminary proposal was probably the most important part of the project.
}{Jordan}

\noindent In this excerpt, Jordan described criteria that could be considered when choosing a project, which is indicative of the \choice\  phase. Specifically, she indicated that it is important to choose a project that aligns with group members' existing skills and potential to \reviewerA{``}expand their knowledge.\reviewerA{''} Additionally, Jordan described learning the importance of balancing work and managing time among team members, which is indicative of her own new \managerial\  knowledge.

When explaining why he chose his sonoluminescence project, Francisco described \domain\  knowledge:

\quotate{
The thought of sound waves in water capturing a micron-sized bubble and causing it, through some mechanism, to glow, to discharge light, it sounded very interesting. Very exciting! Especially when he [the professor] explained that the actual mechanism behind it isn't fully, completely understood even today. I mean, if you look up online, you don't find any clear answers as to what exactly is going on.
\ldots\ The estimated heat range given off by these bubbles I think is, like, thousands of Kelvin. I mean, it's huge. It's hotter than the surface of the sun!}{Francisco}

\noindent In this excerpt, Francisco described his initial reaction to his professor's explanation of the project, which is indicative of the \choice\  phase. Specifically, he described being interested in and excited about the topic, leading him to search for more information online. Francisco learned about sonoluminescence from both his professor and the internet, which is indicative of his own new \domain\ knowledge.

In total, all 15 participants each discussed \understanding\ during the \execution\ phase several times during their interviews. Of these, 8 described gaining knowledge about the \NOS, 9 described gaining \domain\ knowledge, and 15 students described gaining \system\ knowledge. For example, when describing a problem he encountered while working on his project, Carlos said,

\quotate{It was the third trial especially that we were like---took a step back and said, \reviewerA{`}Wait a second. We are not having very repeatable results.' And so, we struggled for a while trying to figure out what that was. We eventually decided to turn off the laser, let it cool, [then] turn it back on. One of our theories was that the heat affected it, and that seemed to do the trick. When we turned it back on, we measured intensities that were similar to our first trial.}{Carlos}

\noindent In this excerpt, Carlos described a repeatability issue with the apparatus, which is indicative of the \execution\  phase. Specifically, Carlos described a troubleshooting process in which his group successfully identified, diagnosed, and resolved a problem. Resolving the problem involved learning that the temperature of the laser impacted its performance, which is indicative of his own new \system\  knowledge.

When recalling a memorable experience while working on his diffraction project, Alyssa described \domain\  knowledge:

\quotate{The parts that I think were memorable were, like, standing in front of the whiteboard, drawing out diagrams, trying to figure out what was happening. Like, are we taking the inverse, are we taking the Fourier transform? What are our parameters?  \ldots\ That's probably the most distinctive memory I had, was us definitely grinding over the theory a bit.}{Alyssa}

\noindent In this excerpt, Alyssa referred to \reviewerA{``}grinding over the theory,\reviewerA{''} or the ongoing process of understanding how to model their project, which is indicative of the \execution\  phase. Specifically, Alyssa described working to understand the theoretical background of his apparatus. This included collaboratively resolving questions about the mathematical aspects of the theory underlying the project, which is indicative of his own new \domain\  knowledge.

Alyssa recounted a second memorable moment, in which he described \NOS\  knowledge:

\quotate{Like I said, it [the project] was this ongoing process, so I was always just making sure, like, \reviewerA{`}Do I really know what's going on?' Just going back to the papers because the code never seemed to work a lot of times during the first couple weeks. So, I guess that's more of a negative---I wouldn't even call it a bad thing to happen. It's part of the process. The ordeal, I believe, helps you learn better.}{Alyssa}

\noindent In this excerpt, Alyssa described creating code to analyze data, which is indicative of the \execution\  phase. Specifically, he described the frustrating process of troubleshooting his code and double-checking his understanding of the analysis methods in reference to existing literature. Alyssa framed \reviewerA{``}the ordeal\reviewerA{''} as part of conducting research, which is indicative of his new \NOS\  knowledge.

In total, four participants each discussed \understanding\ in the \deliverables\ phase once or twice during their interviews. Due to the small number of excerpts, we were unable to identify patterns in the types of knowledge that students described during this phase, so we avoid making fine-grained claims about subthemes in favor of identifying general instances of \understanding. For example, when Olivia explained what she was working on during the last week of her thermal lensing project, she said,

\quotate{We kind of did a few more rounds of data taking for the beam width and beam radius just to make sure that we were understanding the experimental setup. But, other than that, we kind of focused on, \reviewerA{`}Okay, let's get this presentation together. What do we want to present? What do we not want to present?' \ldots\ And kind of trying to come together and was like, \reviewerA{`}What are we actually going to share with the class? What's actually important for understanding this phenomenon and understanding what our experiment was all about?'}{Olivia}

\noindent In this excerpt, Olivia described preparing a presentation, which is indicative of the \deliverables\ phase. Specifically, she described collecting additional data to confirm her group's understanding of their experiment and prioritizing which information to present to her peers, which is indicative of her \understanding.

Our results suggest that managerial and domain knowledge can be developed when choosing or proposing a project, domain knowledge and knowledge about the nature of science can be developed when executing the project, and new knowledge about the project can be generated when creating a final deliverable. Such instances of students developing their own new understanding are interactional elements of the ownership-as-relationship model.

\vspace{-10pt}
\section{Limitations, discussion, and implications}

Although our study was multi-institutional in scope, due to its qualitative nature, we cannot make probabilistically generalizable claims. Rather, we aim for theoretical generalizability~\cite{Eisenhart2009}. By characterizing students' new understanding about a project during different project phases, our work adds nuance and complexity to the model of project ownership presented in Ref.~\cite{Dounas-Frazer2019}. Our findings are limited in part by our inability to make fine-grained claims about knowledge generation during the deliverables phase. In addition, our analysis included only interactions that were also assigned one of the three phases, which means that atypical phase-less interactions were not considered. Despite these limitations, we can nevertheless make a few key claims.

Our work suggests that students' sense of ownership can entail their own new knowledge about the nature of science. When combined with prior work that found a correlation between students' pre-project views about experimental physics and their post-project sense of ownership~\cite{Dounas-Frazer2018a}, this finding suggests a reciprocal relationship between views about the nature of science and sense of ownership: sophisticated views may support students to view setbacks as part of what Alyssa referred to as \reviewerA{``}the ordeal\reviewerA{''} rather than evidence of personal failings; and, like for Alyssa, the process of working on one's own project may support students to develop sophisticated views about science. Such reciprocity may be connected to the coupled shifts in students' self-efficacy and views about the nature of science that sometimes happen when students work on semester-long projects~\cite{Quan2016}. It may also be connected to the reciprocal relationship between ownership and motivation demonstrated by Milner-Bolotin~\cite{Milner-Bolotin2001} and the cyclical relationship between ownership, motivation, and competence demonstrated by Enghag~\cite{Enghag2006}. Future research could explore how interconnections between ownership and views about the nature of science might be mediated by students' self-efficacy or motivation.

Additionally, we have demonstrated that all three project phases present opportunities for students to come to view a project as their own through developing their own new understanding about the project. For instructors who aspire to foster a sense of ownership among students working on experimental projects, it may not be enough to teach about the concepts, principles, and models relevant to the project domain and the equipment, software, and components that comprise students' apparatus. It could be beneficial to also ensure that managerial knowledge and knowledge about the nature of science are explicitly taught and valued throughout the project. Indeed, as Jordan came to realize, students' managerial choices at the start of a project have implications for all subsequent phases.

Comparing the type of ownership that manifests when students work on projects in a lab course to that which manifests when gardeners work on plots in a community garden can provide insight into classroom dynamics and even motivate future research directions. For example, newcomers to gardening need different kinds of support during different seasons because each season requires the gardener to tend their plot in unique ways; nurturing seedlings is different from gauging the ripeness of a tomato. Similarly, physics students need different kinds of support during different project phases. In ongoing work, we are exploring the role of student-instructor interactions in fostering students' sense of ownership during the choice, execution, and deliverables phases. 

\vspace{-8pt}
\acknowledgments{\confidentialreview{S.\ Lavender and L.\ Torres helped with undergraduate research logistics, A.\ Wooley and R.\ Hart helped interpret results, and C.\ Ramey II and A.\ Gupta helped shape our ideas about ownership. This material is based on work supported by the NSF under Grant No.\ 1726045 and by the WA NASA Space Grant Consortium under Grant No.\ NNX15AJ98H.}{Removed for confidential review.}}

\newpage

\bibliography{./perc2021_database} 

\end{document}